\documentclass[journal]{IEEEtran}
\ifCLASSINFOpdf
\else
\fi
\usepackage{graphicx}  
\usepackage{bm}        
\usepackage{amssymb}   
\usepackage{amsmath}
\usepackage{float}
\usepackage{color}
\usepackage{physics}

\begin{document}
%
\title{Passive Optical Phase Stabilization on a Ring Fiber Network}
%
%
%

\author{Liang Hu,~\IEEEmembership{Member,~IEEE,} Xueyang Tian, Long Wang, Guiling Wu,~\IEEEmembership{Member,~IEEE,} and Jianping Chen
\thanks{Manuscript received xxx xxx, xxx; revised xxx xxx, xxx. This work was supported in part by the National Natural Science Foundation of China (NSFC) (61627871, 61535006, 61905143), and in part by science and technology project of State Grid Corporation of China (No. SGSHJX00KXJS1901531). (\textit{Corresponding author: Liang Hu and Guiling Wu})}
\thanks{The authors are with the State Key Laboratory of Advanced Optical Com- munication Systems and Networks, Department of Electronic Engineering, Shanghai Jiao Tong University, Shanghai 200240, China, with Shanghai Institute for Advanced Communication and Data Science, Shanghai Jiao Tong University, Shanghai 200240, and also with Shanghai Key Laboratory of Navigation and Location-Based Services, Shanghai 200240, China (e-mail: liang.hu@sjtu.edu.cn; txy0220@sjtu.edu.cn; wl$_{-}$hit@126.com; wuguiling@sjtu.edu.cn; jpchen62@sjtu.edu.cn)}
}

%
%

\markboth{JOURNAL OF LIGHTWAVE TECHNOLOGY,~Vol.~xxx, No.~xxx, March~2020}%
{Shell \MakeLowercase{\textit{et al.}}: Bare Demo of IEEEtran.cls for IEEE Journals}
%



\maketitle

\begin{abstract}
Optical frequency transfer provides the means for high-fidelity frequency transfer across thousands of kilometers. A critical step in the further developing optical frequency transfer is its capability to transfer a high spectral purity feature from ultrastable lasers  or optical clocks to any remote locations and, at the same time, its adaptability  to incorporate the optical frequency transfer technique into any existing communication networks with different topologies. Here we for the first time report a technique that delivers optical-frequency signals to multiple independent remote hubs along a ring optical-fiber network with passive phase stabilization. The technique automatically corrects optical-fiber length fluctuations of arbitrary hubs along the loop by mixing and shifting optical signals.  Without the help of an active phase tracker and a compensator, it could significantly mitigate some technical problems such as the limited compensation speed and phase recovery time, the phase jitter contamination caused by  the servo bump in conventional phase noise cancellation.  Moreover, by transmitting optical signals along both directions using the same optical source, it can improve the signal-to-noise ratio at each hub. This technique maintains the same delay-limited phase noise correction capability as in conventional techniques and, furthermore, improves the phase jitter by a factor of 3, opening a way to a broad distribution of an ultrastable frequency reference with high spectral purity and enabling a wide range of applications beyond metrology over a ring fiber network with the naturally impressive reliability and scalability.
\end{abstract}


%
\IEEEpeerreviewmaketitle

\section{Introduction}

\IEEEPARstart{P}{recision}  timekeeping is a prerequisite for so many applications, ranging from navigation \cite{ludlow2015optical, riehle2017optical}, communication networks, radio astronomy \cite{he2018long, clivati2017vlbi} to searching for beyond-standard-model physics \cite{van2015search, safronova2018search}. Today’s most precise clocks are optical clocks with trapped atoms or ions, which use the ultrastable lasers to detect the optical frequency of an electron transitioning between two atomic states as the timebase  \cite{PhysRevLett.120.103201, schioppo2017ultrastable, McGrew:2018aa}. The outstanding performance makes the optical clocks and ultrastable lasers become ideal tools for precision measurements and fundamental physics tests, such as general relativity, temporal variation of the fundamental constant \cite{parker2018measurement}, searching for dark matter, chronometric geodesy \cite{grotti2018geodesy}, and gravitational waves \cite{kolkowitz2016gravitational, graham2013new, hu2017atom}. However, these clocks and ultrastable lasers are cumbersome and expensive and only available at national metrology institutes and several universities \cite{ PhysRevLett.120.103201, schioppo2017ultrastable, takamoto2005optical}. This causes a strong motivation to develop effective systems for comparing and distributing these sources of ultraprecise frequency signals. Among them, the fiber-optic frequency dissemination technique has been recognized as an ideal solution for ultra-long haul dissemination because of  fiber-optic's particular advantages of broad bandwidth, low loss, and high immunity to environmental perturbations, etc \cite{ma1994delivering}.  


%


\begin{figure*}[htbp]
\centering
\includegraphics[width=0.9\linewidth]{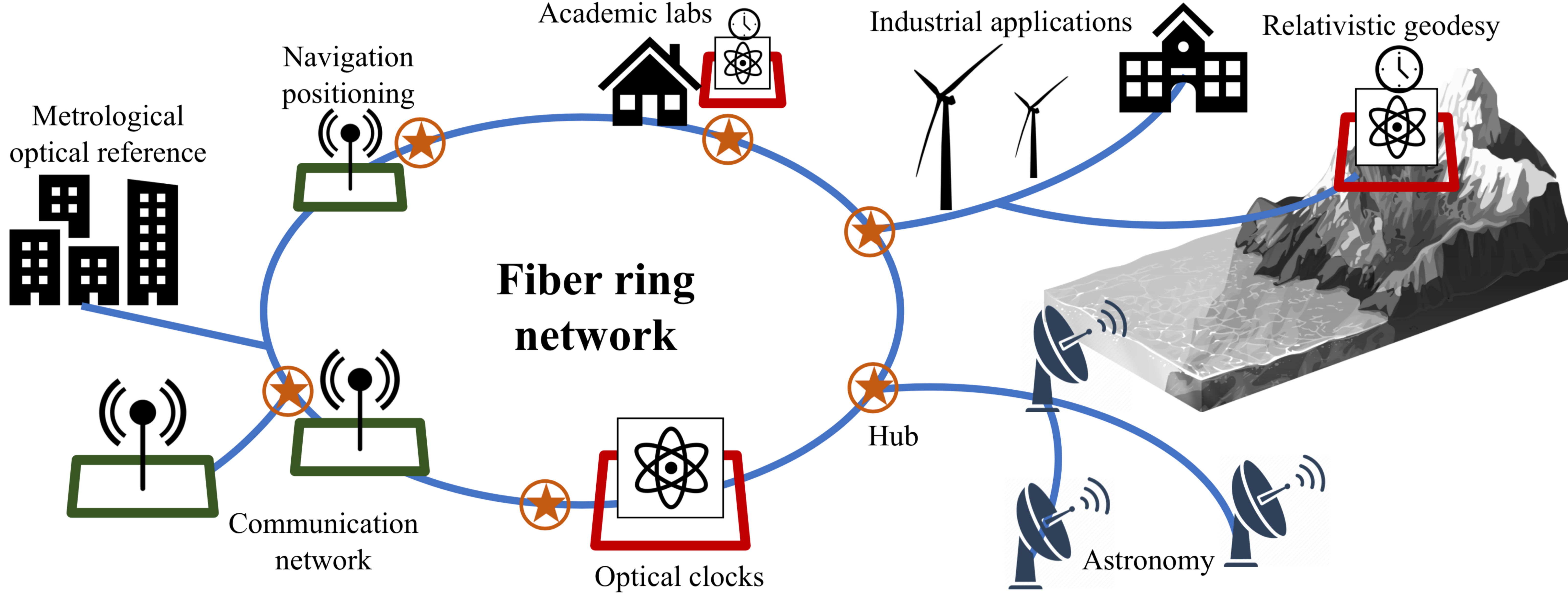}
\caption{A prospect hybrid ring and bus fiber topologies for the dissemination of optical frequency to academic labs,  industrial and scientific applications used for navigation, communication networks, radio astronomy and precise test of relativistic geodesy. A ring is created with some number of optical frequency hubs. Each hub on the ring then acts as the center of the star with multiple point-to-point links emanating and terminating at various remote nodes.}
\label{fig1}
\end{figure*}


Solutions based on fiber transmission have been aiming for suppressing the fiber-induced phase noise to retrieve precise frequency information at remote locations. To achieve this aim, active compensation schemes as first demonstrated in 1994 by Ma \textit{et al.} have been proposed to cancel the fiber-induced phase drift and implement highly stable optical frequency distribution \cite{ma1994delivering, predehl2012920, droste2013optical2, calonico2014high}. It generally utilizes the phase error from a round-trip probe signal to achieve the feedback control of compensators. The compensators mainly include variable delay lines \cite{lopez2010high} and phase-locked loops (PLL) \cite{ma1994delivering}. Although this scheme can accomplish very high phase stability, the response speed and phase recovery time are restricted by the compensators' parameters and optimization. Moreover,  much attention has paid into the relative long-term frequency instability and accuracy, while little into high spectral purity  of the transferred light. The possibility of transferring the spectral purity of an ultrastable laser across different locations is beneficial to the increasing requirement of high frequency stability lasers for optical atomic clocks and high-resolution spectroscopy \cite{coddington2007coherent, argence2015quantum}.   Optical frequency transfer with high spectral purity enables such performances to be copied to any laser in any locations, with a simplification of the experimental setup. This is especially relevant when several ultrastable lasers at different locations  are needed, but only one ultrastable cavity or clock exists.

In order to surmount the above mentioned barriers, {passive phase noise cancellation has drawn extensive attention for fiber-optic radio frequency transfer \cite{pan2016passive, he2013stable}. The passive phase noise cancellation technique can realize rapid and endless phase fluctuation compensation, and also get rid of complicated phase error detection and feedback circuits. However, the passive phase noise cancellation technique used for RF frequency transfer is not directly applicable for fiber-based optical frequency dissemination by multiplying and dividing the frequency of the transferred optical carrier, such as 1550 nm, itself. In our previous work, we have extended the passive phase noise cancellation technique in optical frequency transfer by detecting and compensating optical phase noise with different optical signals along the single path \cite{hu2020passive}}.  The main drawback related to this technique is the different frequencies between the detection and the compensation beam, leading to that two different frequencies will be received at the remote site and, therefore, a narrow bandpass optical filter has to be adopted to remove the undesired signal, which may cause additional decoherence on the transferred light.



Over the last decade, extensions have been proposed that can provide stabilized optical-frequency signals at intermediate sites along the length of optical fiber \cite{grosche2014eavesdropping, bai2013fiber, bercy2014line, schediwy2013high}. However, as phase stabilization at the intermediate sites achieved  by mixing signals received from the source and the far end of the fiber, this approach is limited to fiber links with a bus topology. Moreover, if the stabilization servo of the main link fails, then transfer to all downstream remote sites will cease to be stabilized. To overcome this main drawback, ultrastable optical frequency dissemination schemes on a star topology optical fiber network have been proposed and demonstrated \cite{schediwy2013high, hu2020multi, wu2016coherence}. Using this method, a highly synchronized optical signal itself can be recovered at arbitrary remote locations by actively compensating the phase noise of each fiber link at each user end \cite{schediwy2013high, hu2020multi, wu2016coherence}. However, the maximum node accommodation capability will be limited by the radio frequency  (RF) bandwidth of AOMs to distinguish the optical frequency between the accommodated nodes and the bandwidth of the electrical bandpass filters. Moreover, the existing schemes to support optical communication based on bus and star topologies have limited scalability  and reliability \cite{breuer2011opportunities, effenberger2015pon, chanclou2012network}. On the contrary, because of the self-healing characteristic of the ring network, in particular, the dual-fiber ring, has a natural advantage in the network reliability \cite{gou2018tangent}. Although the number of fibers required in the dual-fiber ring doubles that in the single-fiber ring, the dual-fiber ring network has a protection mechanism and can carry out the protection of multiple faults, resulting in shortening the recovery time  and possessing higher reliability \cite{li2018novel, sun2007single}. In addition, by deploying optical amplifiers in remote nodes, the scale of the ring network can be increased dramatically \cite{zhang2016efficient}. With the continuous extension of the optical frequency transfer network, the reliability and scalability will become more important \cite{effenberger2015pon, chanclou2012network, hu2020multi}. Owing to the prominent advantages, the performance and compatibility of optical frequency transfer on a fiber ring network have to be investigated theoretically and experimentally.

\begin{figure*}[htbp]
\centering
\includegraphics[width=0.95\linewidth]{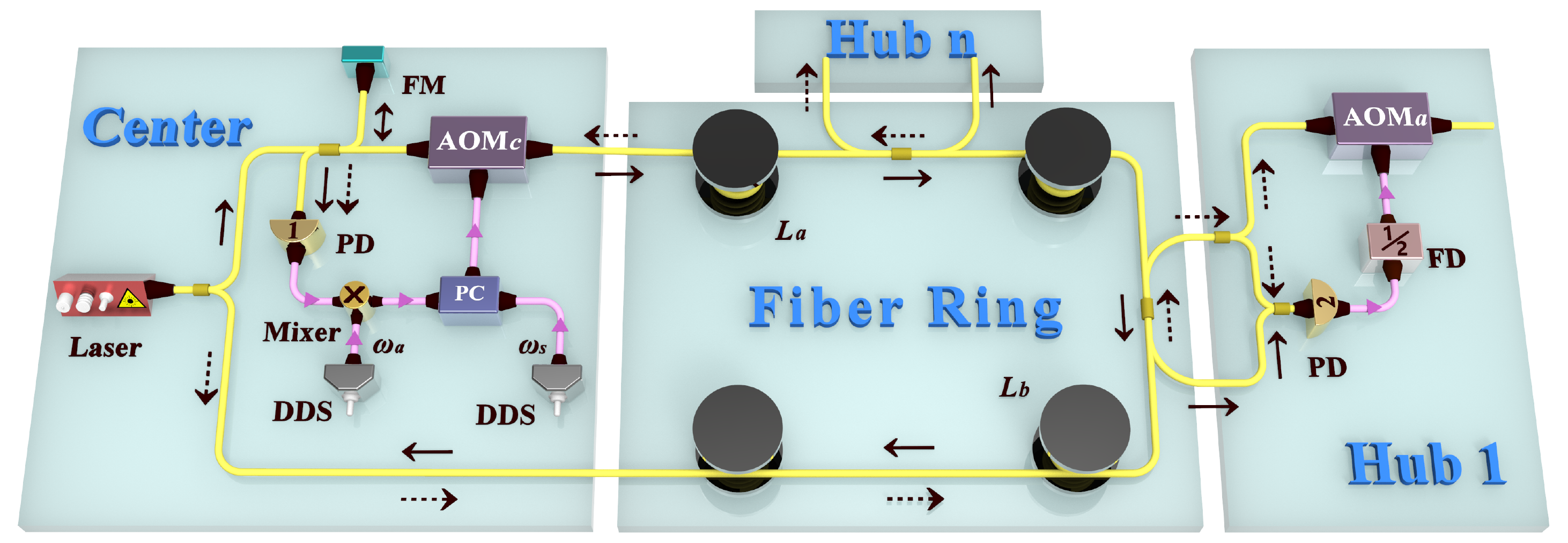}
\caption{Schematic diagram of our optical frequency transfer over a ring fiber network with passive phase stabilization.  We tapped bidirectional lights with the assistance of a $2\times2$ optical coupler at each hub. The optical phase introduced by environment perturbations on the fiber links is passively compensated at  each hub. Electrical bandpass filters are not shown for conciseness. AOM: acousto-optic modulator, FM: Faraday mirror, DDS: direct-digital synthesizer, PD: photo-detector, FD: frequency divider, PC: power combiner. The  solid and dashed arrows represent the light propagation along the clockwise and anticlockwise directions, respectively.}
\label{fig2}
\end{figure*}

In this paper, a passive arbitrary-access stable optical phase delivery scheme based on a ring fiber network is proposed and experimentally demonstrated. In comparison with the previous schemes \cite{ma1994delivering, predehl2012920, droste2013optical2, calonico2014high}, precise phase correction is obtained by embedding the phase information into an RF signal and shifting a copy of the  optical frequency signal with the amount of phase noise introduced by the fiber loop to avoid having to actively stabilize the optical frequency signal.  The scheme we proposed largely simplifies the setup at the central station and the hubs simultaneously, and leaves the hubs to independently control the fiber noise cancellation systems as performed in \cite{schediwy2013high, hu2020multi, wu2016coherence}. Moreover, with the proposed configuration, one of  the directions will only provide one optical signal at each hub's output instead of two optical signals \cite{hu2020passive}.

The proposed technique together with optical frequency transfer over a star topology \cite{schediwy2013high, hu2020multi, wu2016coherence} provides a promising way to implement a robust optical frequency transfer network as illustrated in Fig. \ref{fig1}. Depending on the size and distance of the network, a ring can be created with some number of optical frequency hubs which are all connected together to keep failure rate as low as possible. At the same time, the various hubs on the ring then act as the center of the star with multiple point-to-point links, emanating and terminating at various remote nodes. These individual remote nodes may be subject to failure, so they are generally located at non-critical positions and can accept occasional outages. The ring, on the other hand, keeps the hubs communicating constantly and makes the overwhelming majority of the network fault-free \cite{effenberger2015pon, chanclou2012network, hu2020multi}. This hybrid optical frequency transfer network could be used in probes of fundamental physics and detection of submarine earthquakes by means of deep-sea fiber cables \cite{marra2018ultrastable}, among other applications \cite{parker2018measurement, grotti2018geodesy, kolkowitz2016gravitational, graham2013new, hu2017atom}. At the same time, with the assistance of optical combs, stable and accurate microwave signals can be obtained  and can be used in a variety of areas including communication, navigation, radar, radio astronomy, and fundamental physics research as illustrated in Fig. \ref{fig1}.

The article is organized as follows. We illustrate the concept of coherent optical phase dissemination with passive optical phase stabilization on a ring fiber link in Sec. \ref{sec2} and present in Sec. \ref{sec3} the delay limited phase-noise  power spectral density (PSD). We discuss the experimental set-up and experimental results in Sec. \ref{sec4} and illustrate representative features related to the proposed scheme in V. Furthermore, we briefly present a discussion in Sec. \ref{sec5}.  Finally, we conclude in Sec. \ref{sec6} by summarizing our results.

\section{Concept of optical frequency transfer on a ring fiber network}
\label{sec2}

A schematic diagram of the proposed technique is illustrated in Fig. \ref{fig2}. Here we briefly describe the principle of our optical frequency transfer on a ring fiber link. An optical-frequency signal $\nu$ is divided into 2. The two parts are, respectively, sent from the signal source to the central site along the clockwise and anticlockwise directions over a ring fiber link. The laser frequency $\nu$ propagating clockwise is again split into 2. One part is reflected by a Faraday mirror as a reference signal and the remaining one is downshifted by an angular frequency $\omega_s$ with an acousto-optic modulator (AOM) denoted as AOM$c$. The laser frequency propagating anticlockwise is directly injected into the fiber loop, passes through the fiber loop and then returns back in the AOM$c$ located at the central site. The single-trip signal propagating along the anticlockwise direction is mixed with the input ultrastable laser onto a photodetector 1 (PD1). The beat-note frequency is $\omega_s$, exhibiting the single-trip fiber phase noise, $-\phi_p$. After mixing with an another frequency of $\omega_a$ ($\omega_a>\omega_s$) with the assistance of a frequency mixer, the lower sideband signal is extracted and then applied to the RF port of the AOM$c$ together with $\omega_s$, resulting in a desirable clockwise optical signal with the angular frequency of $\nu-\omega_a+\omega_s$.


Now we consider the extraction of the ultrastable signal along the fiber loop with a $2\times2$  optical coupler, enabling us to extract both the clockwise and anticlockwise signals from the loop fiber link, at a distance $L_a$ from the central site along the clockwise direction and $L_b$ from the central site along the anticlockwise direction with the total fiber link length of $L$ ($L=L_a+L_b$). The anticlockwise signal has a frequency $\nu$ and exhibits the phase fluctuation of $\phi_b$, and the desirable clockwise signal with the angular frequency $\nu-\omega_a+\omega_s$ at arbitrary hubs exhibits the phase fluctuations $-\phi_p+\phi_a=-\phi_b$, where $\phi_a$ and $\phi_b$ are the phase noise of the fiber sections $L_a$ and $L_b$, respectively.  To compensate the phase noise of the anticlockwise wave, we detect the beat-note of the two extracted signals onto the PD2. The beat-note frequency is thus $\omega_a-\omega_s$, exhibiting a phase fluctuation of $2\phi_b$. The signal frequency is divided by 2, filtered, and drives an AOM (AOM$a$, $-1$ order) to correct the phase fluctuations of the extracted anticlockwise  signal. The frequency of the extracted anticlockwise  signal, after passing through the AOM$a$, is thus downshifted to $\nu-0.5(\omega_a-\omega_s)$ and its phase fluctuation is cancelled. With this configuration,  the anticlockwise direction only includes one phase stabilized optical signal. Compared to our previous passive phase noise cancellation schemes \cite{hu2020passive}, this represents another advantage, that is, no optical filters are required to  remove the unwanted optical signal. Similar compensation can be obtained on the extracted clockwise  signal with a positive optical frequency shifter. However, in this case, the clockwise direction signal includes two optical frequencies and needs an optical filter after the AOM$a$ to select a stable optical frequency signal, which could introduce an additional decoherence effect \cite{hu2020passive}. 

We can clearly see that the optical signal received at arbitrary hubs has the same  phase with the standard optical signal at the central station. Therefore, the phase noise of the optical signal is effectively reduced by simply mixing and shifting optical signals.   

\section{Delay-limited phase noise PSD}
\label{sec3}

In Sec. \ref{sec2}, the description does not take the propagation delay of the fiber sections into account. The capability of the phase noise rejection will be limited by the propagation delay as first pointed out by Williams \textit{et al.} \cite{williams2008high}. By adopting the similar procedure performed in \cite{bercy2014line, williams2008high}, we find that the residual phase noise power spectral density (PSD) at an arbitrary hub along the fiber section $L_b$ in terms of the single-pass free-running phase noise PSD, $S_{\text{fiber}}(\omega)$, and the propagation delay of the fiber loop, $\tau_0$, can be calculated as,
\begin{equation}
\begin{split}
S_{E, b-}(\omega)&=F_{E, b-}\frac{(\omega\tau_0)^2}{3}S_{\text{fiber}}(\omega)\\
&=\frac{(\omega\tau_0)^2}{3}\left(1-3\frac{L_b}{L}\frac{\tau_b}{\tau_0}+2\frac{L_b}{L}\left(\frac{\tau_b}{\tau_0}\right)^2\right)S_{\text{fiber}}(\omega).
\end{split}
\label{eq1}
\end{equation}
where $\tau_b$ is the proragation delay along the fiber section $L_b$. This coefficient factor $F_{E, b-}$ is zero for $L_b = L$, then increases up to one at $L_b=0$.  Following the same procedure, if we apply the phase correction for the clockwise light, the residual phase noise PSD at arbitrary hubs can have a form of,
\begin{equation}
\begin{split}
S_{E, a+}(\omega)&=F_{E, a+}\frac{(\omega\tau_0)^2}{3}S_{\text{fiber}}(\omega)\\
&=\frac{(\omega\tau_0)^2}{3}\left(1-3\frac{L_a}{L}\frac{\tau_a}{\tau_0}+2\frac{L_a}{L}\left(\frac{\tau_a}{\tau_0}\right)^2\right)S_{\text{fiber}}(\omega).
\end{split}
\end{equation}
where $\tau_a$ is the proragation delay along the fiber section $L_a$.




\begin{figure*}[htbp]
\centering
\includegraphics[width=0.95\linewidth]{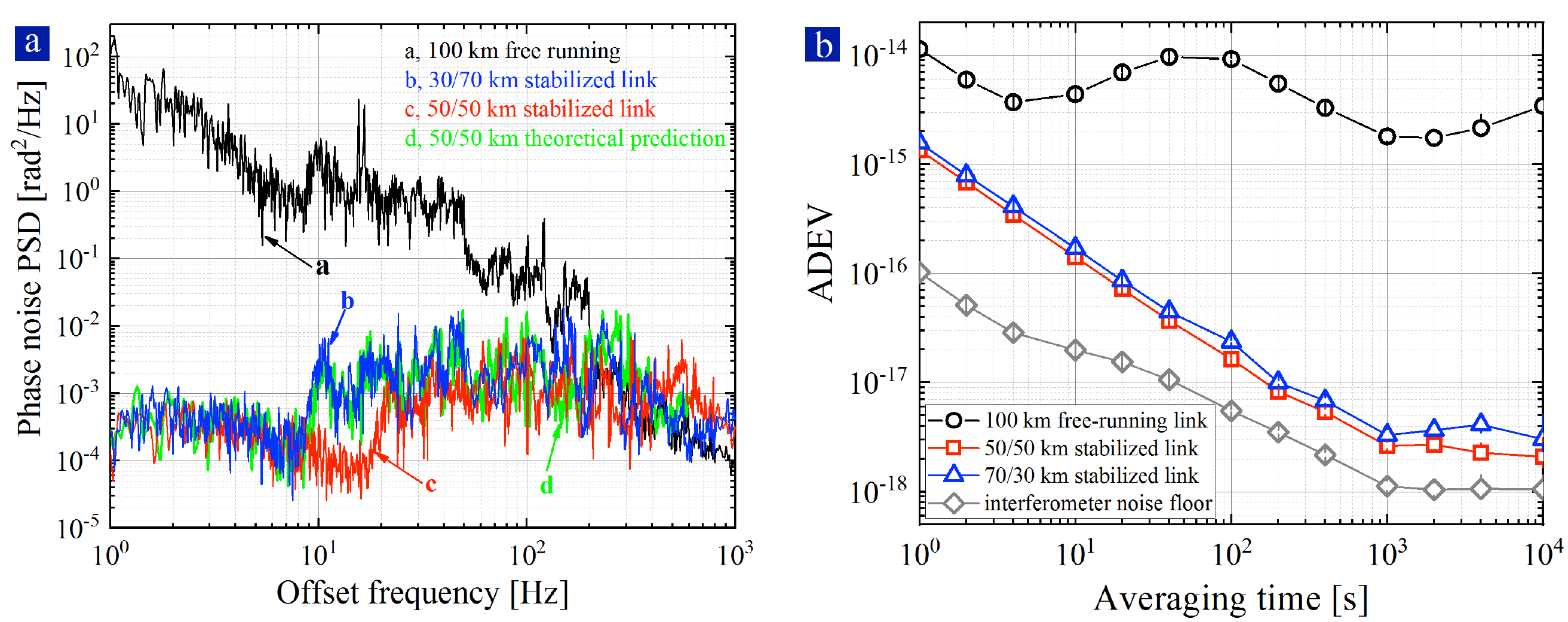}
\caption{(a) Measured phase noise PSDs of the 100km free-running fiber link (black curve) and the stabilized link with passive phase noise cancellation for the 50/50 km hub(\textbf{c}, red curve) and the 30/70 km (\textbf{b}, blue curve) hub. Note that strong servo bumps can be effectively eliminated in the passive phase noise cancellation scheme. The green curve is the theoretical prediction based on Eq. \ref{eq1}. (b) Measured fractional frequency instabilities of the 100 km free-running fiber link (black circles) and the stabilized link for the 50/50 km (red squares) hub and the 30/70 km (blue triangles) hub. The measurement is derived from non-averaging ($\Pi$-type) frequency counters expressed as ADEV. The measured noise floor of the interferometer is also shown (gray diamonds).}
\label{fig3}
\end{figure*}


\section{Experimental apparatus and results}
\label{sec4}

\subsection{Experimental apparatus}
We have demonstrated this technique by using the simplest configuration as shown in Fig. \ref{fig2}. The interferometer is built with fiber optics. The proposed scheme was tested using a narrow-linewidth optical source (NKT X15) at a frequency near 193 THz with a linewidth of 100 Hz. The signal was transmitted along a 100 km fiber link loop. $2\times2$ optical couplers were used to extract both clockwise and anticlockwise light at the most symmetric position, 50/50 km ($L_a/L_b$), and a relative most asymmetric one, 30/70 km, over the 100 km ring fiber link. Here we set $\omega_s=2\pi\times45$ MHz and $\omega_a=2\pi\times80$ MHz. Before dividing the frequency of the beatnote at the hub, we mix the beatnote with an assistant frequency of $115$ MHz, and the lower sideband with a frequency of 80 MHz is extracted. All these RF frequencies are provided by a direct-digital-synthesizer (DDS) generator, phase locked to a 10 MHz rubidium clock. With this configuration,  the AOM$_c$ is simultaneously fed by $35$ MHz and $45$ MHz (downshifted mode), and the AOM$_a$ is working at an angular frequency of 40 MHz (upshifted mode), resulting in an out-of-loop beatnote of $40$ MHz for arbitrary hubs. To avoid the nonlinear effect in the fiber, we keep the optical power into the ring fiber link below 5 dBm for each optical frequency: one for the anticlockwise direction ($\nu$) and two for the clockwise direction ($\nu-2\pi\times35$ MHz and $\nu-2\pi\times45$ MHz). However, in the conventional configuration \cite{bercy2014line}, the light transferred to the remote site will directly return back to the local site, resulting in the power of the returning light of $-15$ dBm at the remote site for the 100 km fiber link  when the injection power is  5 dBm at the local site and fiber loss is $0.2$ dB per kilometer. Consequently, we can obtain the gain of the signal-to-noise ratio of approximately 20 dB without the assistance of optical amplifiers in the proposed scheme.


To effectively measure the transfer stability at each hub, all hubs are co-located at the same optical platform as the signal source. The out-of-loop fiber connections were kept as short as practicable and were thermally and acoustically isolated. We use non-averaging $\Pi$-type frequency counters, which are referenced to the RF frequency source from the DDS at the central site, to record the beating frequency between the fiber input light and the output light. Additionally, to measure the phase noise of the optical carrier frequencies at each hub, we perform the measurement by feeding the heterodyne beat frequency together with a stable RF frequency reference produced by the DDS to a phase detector. The voltage fluctuations at the phase detector output are then measured with a fast Fourier transform (FFT) analyzer to obtain the phase fluctuations.

\subsection{Testing the phase noise rejection on hubs}
To characterize optical transfer over the 100 km ring fiber loop, we measured the phase noise PSDs of  the 50/50 km hub and the 30/70 km hub  for both the stabilized and the unstabilized cases. Typically, the phase noise PSD is usually parametrized as \cite{barnes1971characterization, rutman1978characterization},
\begin{equation}
S_{\phi}(f)=\sum_{\alpha=-2}^{2}h_{\alpha}f^{\alpha-2},
\end{equation}
where $f^{\alpha}$ ($\alpha=-2,-1,0,1$ and 2), reflecting the various contributions of noise in the system (i.e., random walk frequency noise, flicker frequency noise, white frequency noise, flicker phase noise and white phase noise).

The phase noise PSDs of the 50/50 km hub and the 30/70 km hub are plotted in Fig. \ref{fig3}(a). Both hubs are very similar and typical for optical fiber links, with noise of approximately 200 rad$^2$/Hz at 1 Hz and $3\times10^{-2}$ rad$^2$/Hz at 100 Hz with a $h_0f^{-2}$ dependency, indicating that the phase noise of the free-running loop is mainly limited by the flicker phase noise. Both compensated phase noise PSDs are  below $10^{-3}$ rad$^2$/Hz between 1 and 10 Hz with a $h_2f^0$ dependency, illustrating that the loop is mainly constrained by the white phase noise after the phase noise compensation. Noise is corrected up to about 400 Hz, which is compatible with the theoretical bandwidth of 500 Hz given by $1/(4\tau_0)$ with $\tau_0$ being the propagation delay of fiber loop $L=100$ km. This limit is the same for both hubs and is mainly determined by the longest propagation delay $\tau_0$. We checked that the noise floors of both outputs were below these PSDs. The noise rejection of around $2\times10^5$ at 1 Hz is also compatible with the theoretical limit given by  Eq. \ref{eq1} as the green curve shown in Fig. \ref{fig3}(a). This shows that the noise rejection is optimized. 


\subsection{Time-domain characterization}

A time-domain characterization of  the frequency stability in terms of overlapping Allan deviation (ADEV) is shown in Fig. \ref{fig3}(b). In this plot, black circle markers indicate the fractional frequency stability of optical carrier frequency dissemination over the 100 km link when passive phase noise cancellation is not activated. Curves with square  and triangle markers represent the stability of the signals with implementing passive phase noise correction for the $50/50$ km hub and the $30/70$ km hub, respectively.  With the implementation of  fiber noise cancellation at the $50/50$ km ($30/70$ km) hub, optical frequency transfer achieves a fractional frequency stability of $1.6\times10^{-15}$ ($1.4\times10^{-15}$) at the integration time of $1$ s, decreases and reaches a floor of approximately $3.0\times10^{-18}$ ($3.0\times10^{-18}$) at $1,000$ s.


We can clearly see that when the fiber noise cancellation setups are engaged, frequency fluctuations can be effectively suppressed and no longer dominate the instability of the optical signals at both hubs. In our experiment, we observe that the stability of optical frequency dissemination is improved by three orders of magnitude at the integration time of 10,000 s.  Note that the noise correction is very robust and that the set-up can operate several days without any cycle slips.  As a comparison, we measured the floor of optical frequency dissemination by replacing each fiber spool with a 1 m fiber plus a 20-dB attenuator. We can observe that the floor of optical frequency dissemination with a stability of $1.0\times10^{-16}$ at 1 s and $1.1\times10^{-18}$ at 10,000 s is obtained. Consequently, the stabilized link is mainly limited by the noise floor. There are several reasons that lead to the floor in the instability including the noise of the imperfect length adjustment and thermal stabilization in the extraction optical set-up, and the interferometric measurement set-up \cite{foreman2007coherent, predehl2012920, droste2013optical2}. We estimate the path length mismatch up to $ 10$ cm. For typical temperature perturbations due to our air conditioning system, with the temperature fluctuation amplitude $1$ K and cycle 3, 600 s, one expects a bump of the ADEV as high as $4\times10^{-18}$ at approximately 1,800 s \cite{tian2020hybrid}.

As calculated by Eq. \ref{eq1}, the ratio of the stability of the $50/50$ km and $30/70$ km hubs should be $\mathcal{R}=0.64$. In our experiment, we obtain the ratio of $\mathcal{R}=1.4\times10^{-15}/1.6\times10^{-15}=0.87$, which has a large deviation from the theoretical one. We attribute this discrepancy to the phase noise introduced by the hub itself such as the photo-detection process. We have to note that the estimation in Eq. \ref{eq1} acquired by the assumption that  the hub will introduce negligible phase noise. In our system, the  phase noise introduced by the hub itself dominates the total phase noise of the hub at the short  fiber section $L_b$ whereas  the residual phase noise of the fiber link becomes the domination when the fiber section $L_b$ is long enough, enabling that the measured results are consistent with the theoretical one as increase of the fiber section $L_b$.


\subsection{Frequency transfer accuracy}

\begin{figure}[htbp]
\centering
\includegraphics[width=0.95\linewidth]{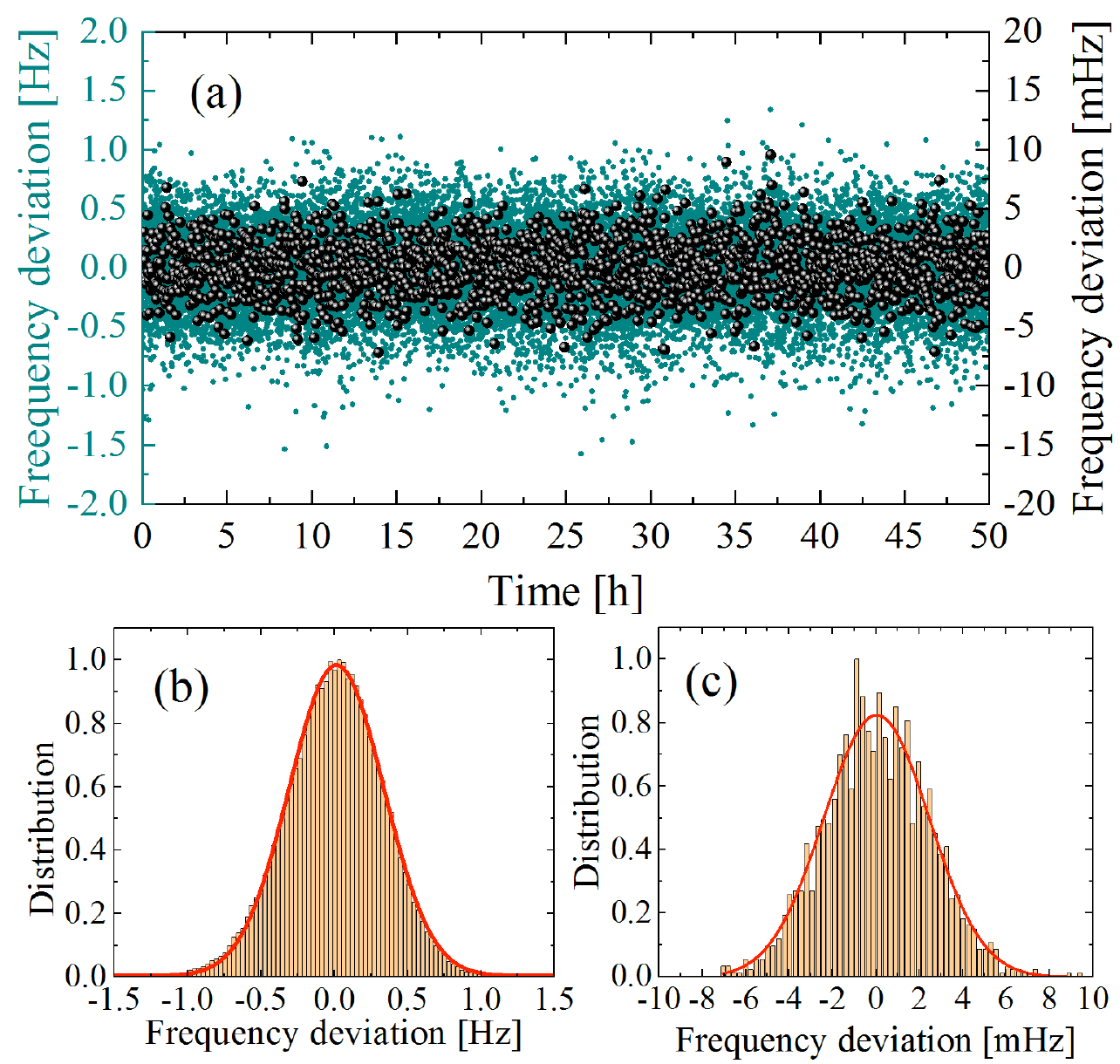}
\caption{ (a) Two-day frequency comparison between sent and transferred frequencies after 50 km over the 100 km ring fiber. Data were taken with dead-time free $\Pi$-type frequency counters with a 1 s gate time (green points, left axis). We calculated unweighted mean ($\Pi$-type) values for all cycle-slip free 100 s long segments, resulting in 1,803 data points (black dots, right frequency axis, enlarged scale). Histograms (brown bars) and Gaussian fits (red curves) for (b) frequency values as taken with $\Pi$-type frequency counters with one second gate time and (c) 1,803 phase coherent 100-second frequency averages with a mean of $2.3\times10^{-18}$ and a standard deviation of $1.2\times10^{-17}$. Taking the long-term stability shown in Fig. \ref{fig3}(b) into account, we determine the statistical uncertainty to be $3\times10^{-18}$.}
\label{fig4}
\end{figure}

We also performed an evaluation of the accuracy of frequency transfer at arbitrary hubs. Figure \ref{fig4} shows the frequency deviation of the beat-note’s data for the 50/50 km hub, recorded with a 1 s gate time and $\Pi$-type counters, over successive 180,300 s (green point, left axis) and the arithmetic mean of all cycle-slip free 100 s intervals (black dots, right axis).  Histograms (brown bars) and Gaussian fits (red curves) of a frequency deviation for the hub after 50 km are also illustrated in Fig. \ref{fig4}(b) and (c). According to the Gaussian fit in Fig. \ref{fig4}(c), the calculated results demonstrate that the mean frequency is shifted by 435 $\mu$Hz ($2.3\times10^{-18}$).  The standard deviation of the 100 s data points is $2.3$ mHz ($1.2\times10^{-17}$) which is a factor of 100 smaller than the ADEV at 1 s as expected for this $\Pi$-type evaluation. Considering the long-term stability of frequency transfer as illustrated  in Fig. \ref{fig3}(b) mainly limited by the flicker frequency noise, we conservatively estimate the accuracy of the transmitted optical signal as shown in the last data point of the ADEV, resulting in a relative frequency accuracy of $3\times10^{-18}$.

\begin{figure*}[htbp]
\centering
\includegraphics[width=0.95\linewidth]{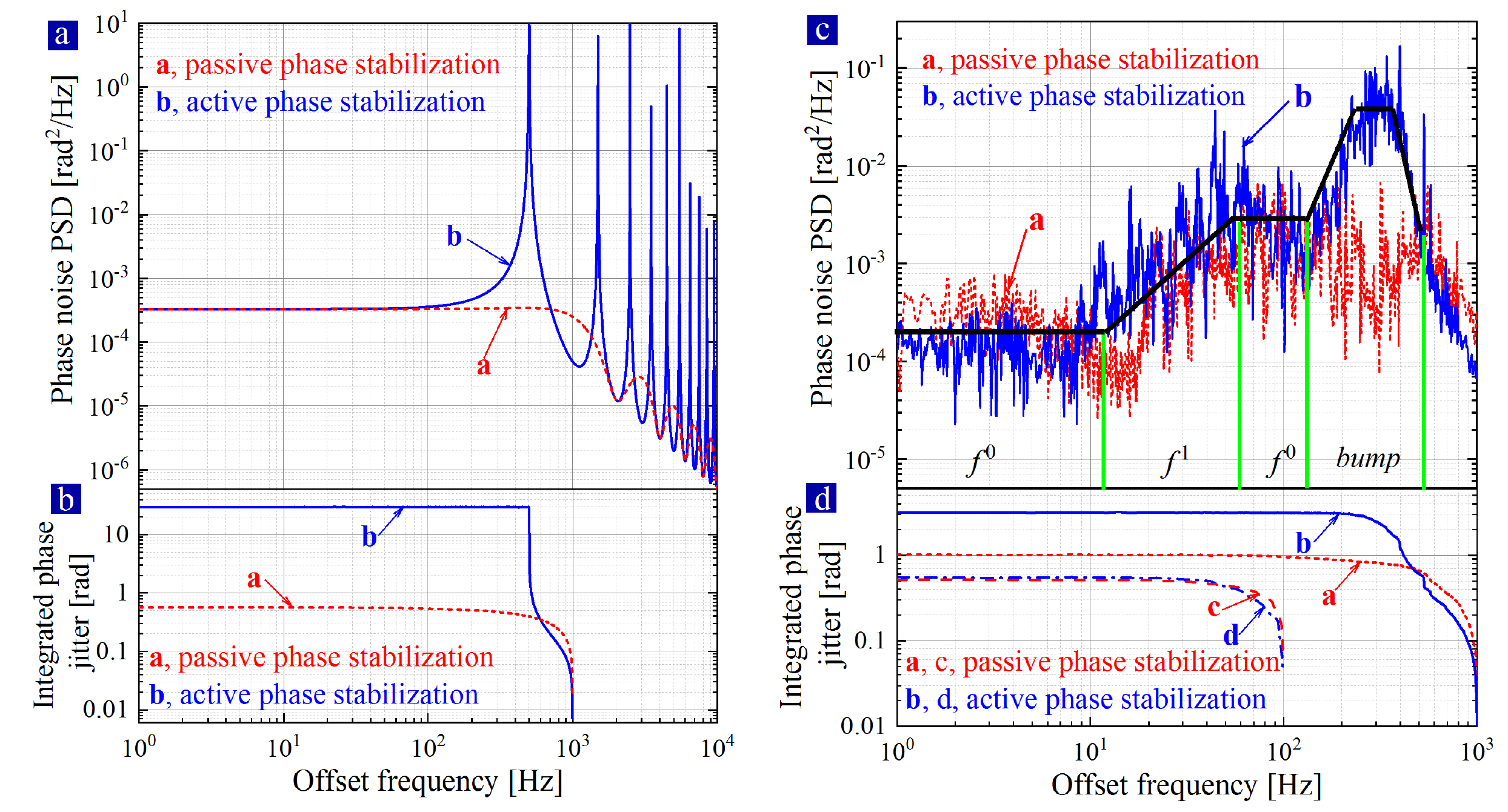}
\caption{(a) Blue solid curve and red dashed curve show, respectively, the residual phase noise PSDs with the active  and passive phase noise cancellation system for  the phase noise PSD of the free-running fiber link $100/f^2$ rad$^2$/Hz.  To maintain a sufficient phase noise rejection capability, the gain has to be tuned large enough, leading to the divergence of the gain amplitude for frequencies equal to integer multiple of $f_0=1/(4\tau_0)$.  (b) The phase jitter integrated from 1 Hz to 1 kHz  for the active (blue solid curve) and passive (red dashed curve) phase noise cancellation system, respectively.  (c) Measured phase noise PSD at the 50/50 km hub over  the 100-km optical link with passive (\textbf{a}, red dashed curve) and active (\textbf{b}, blue solid curve) phase cancellation. Black lines represent the extrapolated noise components. Active phase noise cancellation appears a strong servo bump compared to passive phase cancellation. (d) The phase jitter integrated from 1 Hz to 1 kHz is  $\sim3.2$ rad and  $\sim1.0$ rad for the active (blue solid curve) and passive (red dashed curve) phase noise cancellation system, respectively. As a comparison, the phase jitter integrated from 1 Hz to 100 Hz for the active (\textbf{d}, blue dashed dot  curve) and passive (\textbf{c}, red long dashed curve) phase noise cancellation systems, respectively is also shown.}
\label{fig5}
\end{figure*}


Following the same procedure, the mean frequency offset for the 30/70 km hub was calculated using the total 40,069 $\Pi$-type counter data to be -812 $\mu$Hz ($-4.2\times10^{-18}$) and a standard deviation of the 100 s points is $4.2$ mHz ($2.2\times10^{-17}$). Considering the long-term ADEV at 10,000 s of the data set for the 30/70 km hub of $2.1\times10^{-18}$, we conservatively estimate that the mean frequency offset is $-4.2\times10^{-19}$ with a statistical uncertainty of $2.1\times10^{-18}$ for the 30/70 km hub. We can conclude that there is no systematic frequency shift arising in the extraction setup at a level of a few $10^{-18}$.

\section{Representative features in the proposed technique}
The above section is mainly devoted to characterizing the results of our scheme from the perspective of  conventional optical frequency transfer parameters consisting of the fractional frequency stability, the phase noise PSD and the accuracy as performed in most existing research work \cite{ma1994delivering, predehl2012920, droste2013optical2, calonico2014high}. In this section, we will theoretically study and experimentally demonstrate the representative features of our  proposed scheme, that is, a ring fiber network with passive phase stabilization, including the lower phase jitter and faster phase recovery capability.


\subsection{Lower phase noise and timing jitter}
For active phase noise cancellation system similar with \cite{williams2008high}, the closed-loop transfer function at arbitrary hubs along the anticlockwise direction in the frequency domain can be expressed as,
\begin{equation}
\begin{split}
&H_A(\omega)=F_{E, b-}\int_0^{L}dz\exp({-i\omega(\tau_0+z/c_n)})\\
&\quad\times\bigg[\exp({-i\omega(z/c_n)})-\frac{\cos(\omega\tau_0-\omega z/c_n)}{\cos(\omega\tau_0)}\frac{G(\omega)}{1+G(\omega)}\bigg]
\end{split}
\end{equation}
where $G(\omega)$ is the open-loop transfer function of the compensation system, $L$ is the fiber link length and $c_n$ is the speed of light in the fiber. 

With the same procedure adopted in \cite{williams2008high, bercy2014line}, the transfer function in our passive phase stabilization set-up at arbitrary hubs along the anticlockwise direction can be calculated as,
\begin{equation}
H_P(\omega)=F_{E, b-}\bigg[\frac{3}{2}-\cos(\omega\tau_0)-\text{sinc}(\omega\tau_0)+\frac{1}{2}\text{sinc}(2\omega\tau_0)\bigg]
\end{equation}

Figure \ref{fig5}(a) shows the calculated phase noise PSDs for the stabilized link at the 50/50 km hub by using active (blue solid curve) and passive (red dashed curve) phase noise cancellation system with the phase noise PSD of the 100 km free-running link of $10/f^2$ rad$^2$/Hz. In typical  servo controllers, the gain has to be tuned large enough  to maintain a sufficient phase noise rejection capability. The infinite gain will lead to the divergence of the gain for frequencies equal to integer multiple of $f_0=1/(4\tau_0)=500$ Hz. Here the servo bandwidth is mainly limited by the total fiber length instead of the fiber sections ($L_a$ and $L_b$). It is interesting to note this issue is automatically disappeared in the passive phase stabilization set-up with the optimized gain. To calculate the ratio of the phase jitter between the active and passive phase noise cancellation technique, we integrate the phase noise from 1 Hz to 1 kHz as shown in Fig. \ref{fig5}(b). We can see that more than one order of magnitude of the reduction of the phase jitter can be achieved  for the proposed phase noise cancellation technique.  Note that the integration results for the active phase noise PSD are dependent on the frequency resolution of the simulation. Here the frequency resolution is 1 Hz and the phase jitter will increase more as improving the frequency resolution due to the diverged bump effect.

To experimentally verify the calculated results,  we used the set-up shown in Fig. \ref{fig2} as the passive phase noise system. The active phase noise system we used is similar with our previous multiple-access  optical frequency transfer system \cite{hu2020fundamental}. Figure \ref{fig5}(c) shows the residual phase noise PSDs at the $50/50$ km hub over the 100 km fiber link with passive (\textbf{a}, red dashed curve) and active (\textbf{b}, blue solid curve) phase cancellation. In active phase noise cancellation,  the residual phase noise is essentially limited by the residual fiber noise in the range from 1 Hz to $\sim200$ Hz, with a strong bump appearing significantly at 300 Hz.  The shifted bump position from $f_0=500$ Hz could be from the insufficient gain in the servo controller. On the contrary, the spectral analysis does not report any strong noise contribution in the 300 Hz range with passive phase noise cancellation, allowing that  the bump does not play a role in our passive optical phase noise cancellation concept. The total integrated phase noise (1 Hz to 1 kHz) of the data in Fig. \ref{fig5}(d) for active (blue solid curve) and passive (red dashed curve) phase noise cancellation are 3.2 rad and 1.0 rad, which corresponds to temporal jitters of $\sim$ 2.6 $f$s and 825 as, respectively, enabling the reduction of the phase jitter by a factor of about 3 by adopting passive phase stabilization. As a comparison,  the phase jitter integrated from 1 Hz to 100 Hz is almost identical for both cases as shown in Fig. \ref{fig5}(d). The main bottleneck of our detection scheme is the round-trip propagation delay, limiting the servo bandwidth. This can be solved by dividing the fiber link into several sub-links which could  serve to further  reduce the round-trip propagation delay,  resulting in the improvement of the signal-to-noise ratio in our scheme \cite{lopez2010cascaded}.



\subsection{Faster response speed  and phase recovery time}

\begin{figure}[htbp]
\centering
\includegraphics[width=0.95\linewidth]{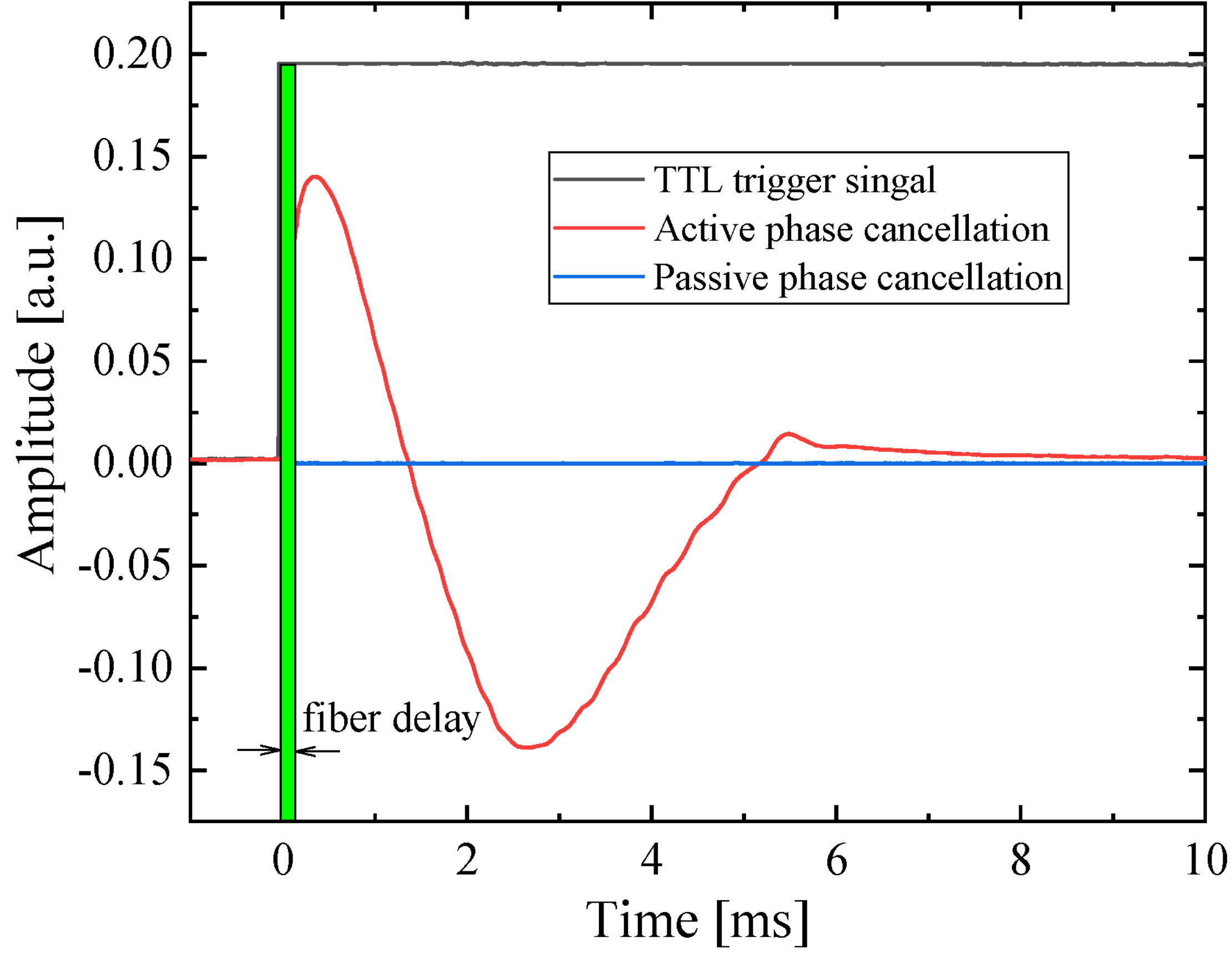}
\caption{Phase recovery behaviour of the 20 km optical path length stabilization with active (red curve) and passive (blue curve) phase correction, respectively. A delay is introduced between the TTL signal (black curve) for switching the light on at 0 s and the activation of the phase stabilization at  $4\tau_0\simeq400$ $\mu$s as indicated by  the shaded green area. }
\label{fig6}
\end{figure}

To examine the characterization of the faster response speed and phase recovery time, we compared two kinds of optical frequency transfer schemes described above  over a 20 km fiber link as performed in \cite{hu2020passive}. To simulate the interruption, we inset one more AOM just after the laser source to switch  the light on/off. The RF port of  the AOM is controlled by a TTL signal which has a rising time of $\sim20$ ns, which can be neglected. Both systems' output was analyzed based on the voltage generated by mixing down the out-of-loop beat to the dc.  Figure \ref{fig6} illustrates the phase recovery time of 20 km optical path length stabilization with active and passive phase correction. We observed that the phase recovery time of optical path length stabilization with active phase noise cancellation has a few strongly damped oscillations of the phase lasting approximately $\sim5$ ms, whereas this time is negligible for our proposed passive phase noise cancellation.  This feature is very beneficial for the case in which the interruptions happen frequently on the long fiber links \cite{predehl2012920, droste2013optical2, calonico2014high}.

\section{Discussion}
\label{sec5}


The above analysis has ignored the effect of the backscattering noise on the frequency transfer performance.  Small-scale inhomogeneities of the refractive index in the fiber cause Rayleigh scattering of the transferring waves. In our case, the backscattered clockwise wave   returns to the access hub and is superimposed upon the extracted anticlockwise wave. Similarly, the backscattered anticlockwise wave returns to the access hub and is superimposed upon the extracted clockwise wave. Consequently, the Rayleigh scattering effect can not be completely avoided in our application. According to the results presented in \cite{hu2020fundamental}, the Rayleigh backscattering induced fractional frequency instability can be as low as a few $10^{-16}/\tau$ ($\tau$ being the averaging time) over a 100 km fiber link. Thus, this effect can be neglected at our precision. 

Our dissemination loop can support multiple hubs simultaneously. Although there is an insertion loss at every hub, proper optical amplifiers such as erbium-doped-fiber-amplifiers (EDFA)  and electrical amplifiers can be used to amplify the desired optical signals and detected RF signals. Thus, it ensures that multiple hubs can be mapped properly along the optical loop link.  Though $N$ copies of hardware for frequency recovery are needed if $N$ hubs are required, all of these copies have the same configuration including fixed optical and electronic components with no tunable parts. It should be noted that several intermediate hubs along the fiber loop were tested besides the 50/50 km and 30/70 km hubs. Because of the similarity among the test results of different hubs, we just show the test results of the two representative hubs selected in the loop fiber link, the most symmetric one (50/50 km) and a relative most asymmetric one (30/70 km).




\section{Conclusion}
\label{sec6}

In conclusion, we demonstrated a technique for dissemination of high-precision optical-frequency signals to multiple independent hubs on a ring optical-fiber network. The technique automatically corrects optical-fiber length fluctuations of each hub along the loop. At the same time, using the same optical source propagating  along both directions can significantly improve the signal-to-noise ratio. The results demonstrate relative frequency instabilities, expressed as overlapping Allan deviation of $1.6\times10^{-15}$ at 1 s averaging time, scaling down to $3.3\times10^{-18}$ at 1,000 s with a $\tau^{-1}$ dependency at the intermediate hub over a 100 km fiber ring. A similar performance is also demonstrated at another hub.  We find no systematic offset between the sent and transferred frequencies within the statistical uncertainty of about $3\times10^{-18}$.


This technique with passive phase compensation maintains the same phase noise rejection capability as in conventional techniques and significantly shortens  the response speed and phase recovery time of optical frequency dissemination and reduces the phase jitter by a factor of 3 compared to the conventional technique, opening a way to a broad distribution of an ultrastable frequency reference with high spectral purity and enabling a wide range of applications beyond metrology over reliable and scalable ring fiber networks.


\ifCLASSOPTIONcaptionsoff
  \newpage
\fi



%

\begin{IEEEbiographynophoto}{Liang Hu}
received the B.S. degree from Hangzhou Dianzi University, China, in 2011, and the M.S. degree from Shanghai Jiao Tong University, China, in 2014. He received the Ph.D. degree from University of Florence, Italy, in 2017 during which he was a Marie-Curie Early Stage Researcher at FACT project. He is currently a Tenure-Track Assistant Professor in the State Key Laboratory of Advanced Optical Communication Systems and Networks, Department of Electronic Engineering, Shanghai Jiao Tong University, China. His current research interests include photonic signal transmission and atom interferometry.
\end{IEEEbiographynophoto}

\begin{IEEEbiographynophoto}{Xueyang Tian}
received the B.S. degree from Shanghai Dianji University, China, in 2017. She is currently a graduate student in the State Key Laboratory of Advanced Optical Communication Systems and Networks, Department of Electronic Engineering, Shanghai Jiao Tong University, China. Her current research interests include photonic signal transmission.
\end{IEEEbiographynophoto}

\begin{IEEEbiographynophoto}{Long Wang }
received the B.S. and M.S. degrees from Harbin Institute of Technology, China, in 2017 and 2019, respectively. He has been admitted as a doctoral student in the State Key Laboratory of Advanced Optical Communication Systems and Networks, Department of Electronic Engineering, Shanghai Jiao Tong University, China. His current research interests include photonic signal transmission.
\end{IEEEbiographynophoto}

\begin{IEEEbiographynophoto}{Guiling Wu}
received the B.S. degree from Haer Bing Institute of Technology, China, in 1995, and the M.S. and Ph.D. degrees from Huazhong University of Science and Technology, China, in 1998 and 2001, respectively. He is currently a Professor in the State Key Laboratory of Advanced Optical Communication Systems and Networks, Department of Electronic Engineering, Shanghai Jiao Tong University, China. His current research interests include photonic signal processing and transmission.
\end{IEEEbiographynophoto}

\begin{IEEEbiographynophoto}{Jianping Chen}
received the B.S. degree from Zhejiang University, China, in 1983, and the M.S. and Ph.D. degrees from Shanghai Jiao Tong University, China, in 1986 and 1992, respectively. He is currently a Professor in the State Key Laboratory of Advanced Optical Communication Systems and Networks, Department of Electronic Engineering, Shanghai Jiao Tong University. His main research interests include opto-electronic devices and integration, photonic signal processing, and system applications. He is a Principal Scientist of National Basic Research Program of China (also known as 973 Program).
\end{IEEEbiographynophoto}

\end{document}